\documentclass[aps,prl,superscriptaddress,twocolumn,showpacs,letterpaper]{revtex4-1}
\usepackage[pdftex]{graphicx}
\usepackage{amssymb}
\usepackage{amsmath}
\usepackage{color}

\begin{document}

\title{Oscillatory coupling between a monolithic whispering-gallery resonator and a buried bus waveguide}

\author{M. Ghulinyan}
\email[]{ghulinyan@fbk.eu}
\affiliation{Advanced Photonics \& Photovoltaics Unit, Fondazione Bruno Kessler,  I-38123 Povo, Italy}
\author{R. Guider}
\email[]{currently at Johannes Kepler Universitat, 4040 Linz, Austria}
\affiliation{Nanoscience Lab., Department of Physics, University of Trento,  I-38123 Povo, Italy}
\author{I. Carusotto}
\affiliation{INO-CNR BEC Center and Department of Physics, University of Trento, I-38123 Povo, Italy}
\author{A. Pitanti}
\affiliation{NEST, Scuola Normale Superiore and Istituto Nanoscienze-CNR, piazza S. Silvestro 12, 56127 Pisa, Italy}
\author{G. Pucker}
\affiliation{Advanced Photonics \& Photovoltaics Unit, Fondazione Bruno Kessler,  I-38123 Povo, Italy}
\author{L. Pavesi}
\affiliation{Nanoscience Lab., Department of Physics, University of Trento,  I-38123 Povo, Italy}


\begin{abstract}
We report on a combined theoretical and experimental study of the optical coupling between a microdisk resonator and a waveguide laying on different planes. While the lateral coupling between a planar resonator and a waveguide is characterized by a unique distance at which the resonant waveguide transmission vanishes because of destructive interference, the vertical coupling geometry exhibits an oscillatory behavior in the coupling amplitude as a function of the vertical gap. This effect manifests experimentally as oscillations in both the waveguide transmission and the mode quality factor. An analytical description based on coupled-mode theory and a two-port beam-splitter model of the waveguide-resonator coupling is developed, which compares successfully both to experimental data and numerical simulations.
\end{abstract}


\maketitle
In typical experiments with integrated circular-shaped resonators, known as Whispering-gallery mode (WGM) resonators, the cavity modes are probed by measuring the transmission spectrum of the bus waveguide \cite{cuttedge}. The resonator-waveguide coupling (RWC) takes place via the overlap of the evanescent fields of the waveguide and resonator optical modes. Therefore, the waveguide transmission spectrum shows a series of transmission dips at the cavity mode frequencies. In a planar geometry, i.e. when the waveguide and the resonator lay on the same plane, the mode coupling is known to grow exponentially when the gap $x$ between these two is decreased. The parameters that are extracted from the waveguide transmission spectra are the transmission value at the bottom of each modal dip, $T_{\rm dip}$, and its linewidth, $\Gamma$.
The exponential variation of the coupling strength as a function of the gap $x$ in this lateral coupling geometry results into clear trends in both $T_{\rm dip}$ and $\Gamma$. In particular,  $T_{\rm dip}$ shows a single minimum at a critical gap $x^*$ at which the field propagating in the waveguide and the secondary emission by the resonator cancel each other on exact resonance due to a complete destructive interference \cite{yariv,spillane}. For larger or smaller gaps, the transmission dips are shallower. On the other hand, following the exponentially growing RWC, the cavity mode linewidth $\Gamma$ increases monotonically as the gap is reduced.

So far, the optical coupling between a single-mode, rectilinear waveguide and a resonator has been usually described in terms of a spatially localized coupling at the minimum separation point. This provides an immediate access to the entire transmission spectrum, including key parameters such as the $T_{\rm dip}$ and the mode quality factor, $Q$, at the  frequency $\omega_0$ ($Q=\omega_0/\Gamma$). The validity of this model in the lateral coupling geometry has proved itself in realizing in-plane all-integrated WGM resonator-strip waveguide systems with an optimized lateral gap that provides an efficient cross-coupling \cite{integr1,integr2,integr3,integr4}.

Recent experimental developments have shown that opting for a \emph{vertical} coupling geometry for the RWC (the resonator and the buried bus waveguide lay on different planes) offers several advantages with respect to the lateral coupling geometry \cite{i3eptl}: it does not require expensive lithographic techniques for gap definition and allows for an independent choice of materials and thicknesses for the optical components. While several experimental realizations of vertically coupled resonator/waveguide systems have been reported~\cite{i3eptl,vert1,ghent1,ghent2}, no specific study of the influence of the vertical gap on the coupling efficiency has been performed yet. The issue of the vertical RWC has been discussed in Ref.~\cite{mandorlo1}, where, the coupling efficiency optimization as a function of lateral (\emph{in-plane}) alignment under fixed vertical gap conditions was addressed.

In this Letter we show that, differently from the in-plane geometry, the critical coupling condition between a resonator and a bus waveguide in a vertical coupling geometry is fulfilled at several values of the gap distance $x$. This unexpected finding results from the spatially extended nature of the vertical RWC; the effective gap between the resonator and the waveguide remains almost constant for a sizeable distance $\Lambda_{fz}$ along the waveguide. Within this \emph{flat zone}, the system can be modeled as a pair of coupled parallel waveguides. By using the coupled mode theory, we find that the interplay between the length $\Lambda_{fz}$ of the flat zone and the coupling length of parallel waveguides, $\Lambda_c$, results in an oscillating effective coupling as a function of the gap $x$. This coupling is characterized by alternating regions of undercoupling and overcoupling separated by different critical coupling points. Our findings are in good agreement with experimental results and are further confirmed through full three dimensional (3D) numerical simulations.

\begin{figure}[t!]
  \centering
 \includegraphics[width=6cm]{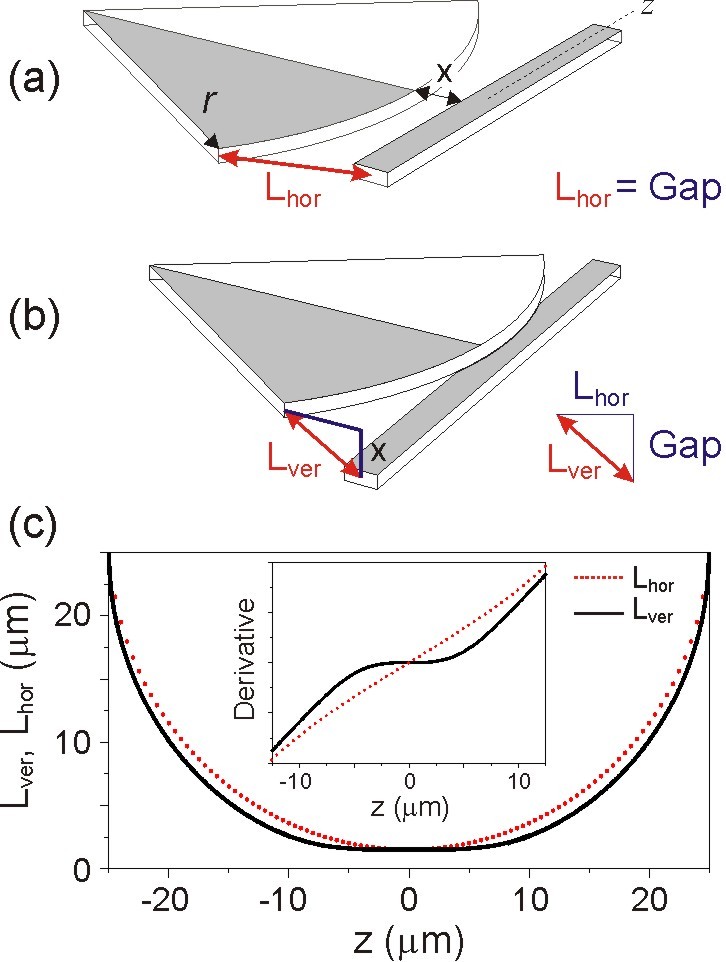}
 \caption{(Color online) 3D sketches of the geometry of (a) in-plane and (b) vertical coupled systems. In the in-plane case (a), the mode-to-mode separation (MM) corresponds to the distance $L_{hor}$ between the resonator and the waveguide. In the vertical case, MM is the distance $L_{ver}$ which is the hypotenuse of the triangle formed by the two catheti, $L_{hor}$ and $x$. As discussed in the text, the gap $x$ is the distance between the two planes where the resonator and the waveguide lay. (c) The MM as a function of $z$-coordinate along the waveguide for the in-plane (dotted line, $L_{hor}$) and the vertical (full line, $L_{ver}$) geometries. In the vertical geometry, $L_{ver}$ exhibits an extended {\em flat zone} in the proximity of the minimum gap. The marked difference between the behavior of the first derivatives $d L_{hor,ver}/d z$ in the two cases is shown in the inset. The resonator diameter is $50~\mu$m. }
\label{m2m}
\end{figure}

First, we start analyzing the geometry of an in-plane coupled system where, for simplicity, only the propagation trajectories of the modes in an infinite strip waveguide and a WGM resonator, separated by a minimum gap of $x$ are considered (Fig.~\ref{m2m}(a)). In this picture, the modes in the waveguide and the resonator propagate along the $z$--axis and a circle of a radius $r$, respectively. Then, the physical distance between the waveguide and the resonator (equivalent to a mode-to-mode separation) at a coordinate $z$ ($-r<z<r$) is given by $L_{hor}=x+r-\sqrt{r^2-z^2}$.

In a vertical coupling scheme, one deals with a 3D picture where the resonator and the waveguide modes lay on different planes, separated by a vertical distance $x$ (Fig.~\ref{m2m}(b)). Considering the simple case, when the resonator is vertically aligned to the $z$--axis at $z=0$ point, the mode-to-mode distance $L_{ver}$ reads as
\begin{equation}\label{lver}
L_{ver}=\sqrt{\left(r-\sqrt{(r^2-z^2)}\right)^2+x^2}.
\end{equation}
At first sight, $L_{hor}$ and $L_{ver}$ do not seem to be particularly different. However, as Fig.~\ref{m2m}(c) shows, $L_{ver}(z)$ differs radically from $L_{hor}(z)$, as the $z$ runs from zero to $\pm r$. In fact, $L_{ver}(z)$ is almost constant over a certain region around $z=0$, which means that the variation in the waveguide-resonator separation here is insignificant. The extension $\Lambda_{fz}$ of the flat zone can be estimated from the inflection points of the second derivative $d^2 L_{ver}/dz^2$, as shown in the inset of Fig.~\ref{m2m}(c); as expected, $\Lambda_{fz}$ depends smoothly on $x$ and is non-zero unless the vertical distance $x$ vanishes.

An important consequence of the existence of the flat zone is that the vertical RWC can be approximated to a coupling between two parallel waveguides within a good accuracy. This allows to employ the coupled-mode theory \cite{cmth1,cmth2} to estimate the coupling coefficients from the dimensions and material refractive indices of the devices used in actual experiments \cite{param}.

\begin{figure} [t!]
\centering
\includegraphics[width=8.5cm]{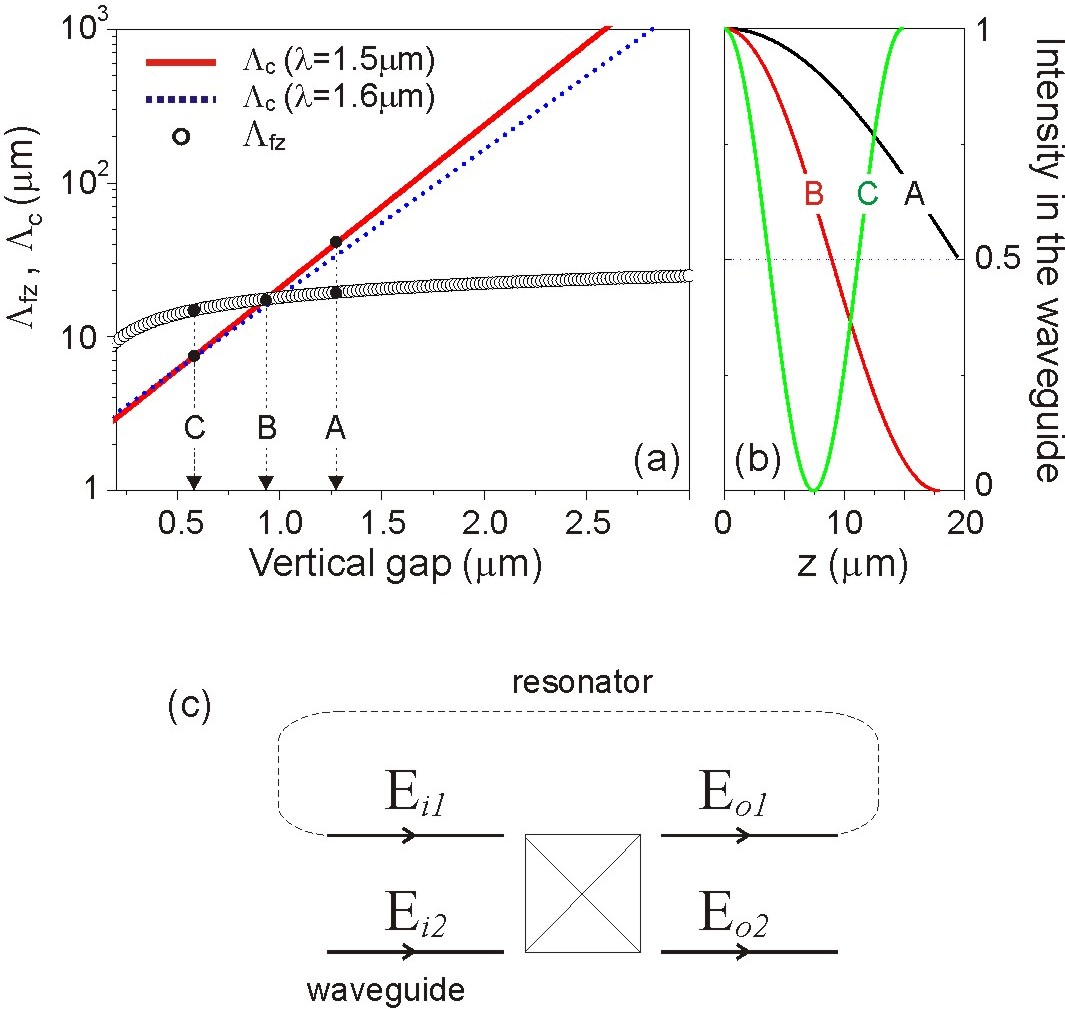}
\caption{
(Color online) (a) The curves for the flat zone extension ($\Lambda_{fz}$, circles) and the coupling length ($\Lambda_c$, line) between parallel waveguides at $\lambda= 1.6 \mu$m show an intersection at a specific coupling gap. An important physical implication of the intersection is that critical coupling between the resonator and the bus waveguide can manifest multiple times at different coupling gaps. A, B and C refer to the situations $\Lambda_{fz}=\Lambda_{c}/2$, $\Lambda_{fz}=\Lambda_{c}$ and $\Lambda_{fz}=2\Lambda_{c}$, respectively. (b) Coupled mode prediction $\cos^2[{\pi z}/({2 \Lambda_c})]$ for the field intensity in the flat zone $z\in[0,\Lambda_{fz}]$ in the three cases A, B and C. (c) Schematics of the vertical RWC reduced to an in-plane coupling model of a two-port beam-splitter.}\label{cpl}
\end{figure}

Figure~\ref{cpl} compares $\Lambda_{fz}$ (for a 50$\mu$m diameter resonator), and $\Lambda_{c}$ (of two parallel waveguides) at a wavelength of $\lambda=1.6~\mu$m. The intersection of $\Lambda_{fz}(x)$ and $\Lambda_{c}(x)$ has important physical implications. While $\Lambda_{fz}$ has a geometrical origin (resonator radius, vertical gap), the coupling length $\Lambda_{c}$ depends on both the waveguide geometry and the refractive indices of the waveguides and the coupling medium. Depending of these, the exponential variation of $\Lambda_{c}(x)$ and the slowly varying $\Lambda_{fz}(x)$ across the crossing can lead to situations where:
\begin{itemize}
  \item [(A)] $\Lambda_{fz}=\Lambda_{c}/2$, which implies that half of the optical power is transferred from the waveguide to the resonator according to $\cos^2[{\pi z}/({2 \Lambda_c})]$,
  \item [(B)] $\Lambda_{fz}=\Lambda_{c}$, which implies that all the optical power is transferred from the waveguide to the resonator,
  \item [(C)] $\Lambda_{fz}=2\Lambda_{c}$, which implies that all the optical power is coupled back into the waveguide by the end of the flat zone extension (Fig.~\ref{cpl}(a),(b)).
\end{itemize}

Based on these results, we proceed with modeling the amplitude of the transmitted light through the straight waveguide coupled to the WGM resonator as a function of frequency $\omega$. According to the schematics shown in Fig.~\ref{cpl}(c), the resonator/waveguide coupling is modeled as a two-port beam splitter. For the output fields $ E_{o1}$, $ E_{o2}$ one has
\begin{subequations}
\begin{align}
        E_{o2}&=\left[E_{i2} \cos\left(\frac{\pi \Lambda_{fz}}{2 \Lambda_c}\right)+\jmath E_{i1} \sin\left(\frac{\pi \Lambda_{fz}}{2 \Lambda_c}\right)\right] e^{\jmath\frac{\omega n}{c}\Lambda_{fz}},\\
         E_{o1}&=\left[E_{i1} \cos\left(\frac{\pi \Lambda_{fz}}{2 \Lambda_c}\right)+\jmath E_{i2} \sin\left(\frac{\pi \Lambda_{fz}}{2 \Lambda_c}\right)\right] e^{\jmath\frac{\omega n}{c}\Lambda_{fz}}.
\end{align}
\end{subequations}
The phase shift and the attenuation of the field after propagation along the rest of the cavity circumference is described by $E_{i1}=E_{o1}\exp[\jmath\frac{\omega n}{c}(L_{cav}-\Lambda_{fz})]$. The refractive index of the resonator is considered to have the form $n=n_0+\jmath\frac{\alpha\lambda}{4\pi}$, where $n_0$ is the real effective refractive index, $\alpha$ is the overall intrinsic loss (material absorption, radiative and surface scattering) of the cavity with a circumference of $L_{cav}=2\pi r$ and $\lambda=2\pi c/\omega$ is the wavelength.

With the definition of two parameters $\theta=\frac{\pi \Lambda_{fz}}{2 \Lambda_c}$ and $\phi=\frac{\omega n}{c}L_{cav}$, the waveguide input and output fields are related as
\begin{equation}\label{tr}
        E_{o2}=E_{i2}\left(\frac{e^{\jmath\phi} \cos\theta-1}{e^{\jmath\phi}-\cos\theta}\right) e^{\jmath\frac{\omega n}{c}\Lambda_{fz}},
\end{equation}
\noindent and, thus, the waveguide transmission is $T(\omega)=|E_{o2}/E_{i2}|^2$.

After some algebraic transformation to Eq.~\ref{tr}, it is possible to rewrite the waveguide transmission in the form
\begin{equation}\label{tr1}
T(\delta\omega)=\left|a+\frac{\jmath b}{\delta\omega-\jmath\Gamma/2}\right|^2,
\end{equation}
\noindent where $\delta\omega=\omega_0-\omega$, $a$=$\cos\theta$,  $b=-\frac{c}{n_0 L_{cav}} e^{-\frac{\alpha L_{cav}}{2}}\sin^2\theta$ and
\begin{equation}\label{gamma}
  \Gamma=\frac{2c}{n_0 L_{cav}}\left(1-\left|\cos\left(\frac{\pi \Lambda_{fz}}{2 \Lambda_c}\right)\right|\times e^{-\frac{\alpha L_{cav}}{2}}\right)
\end{equation}

From Eq.~\ref{tr1}, the typical Lorentzian lineshape of the cavity resonance, centered at $\omega_0$ and having a loss rate of $\Gamma$, can be recognized.  In a weak coupling regime, when $\theta\ll1$, $\Gamma$ tends to its lowest value $\Gamma_0$. The associated intrinsic (unloaded) Q-factor of the cavity is therefore $Q_0=\frac{\omega_0 n_0}{\alpha c}$ (for relatively weak intrinsic loss, $\alpha r\ll 1$). Equations (\ref{tr1}) and (\ref{gamma}) provide close expressions for the resonant waveguide transmission, $T(\omega_0,x)$ and the Q-factor, $Q(\omega_0,x)$, as function of the vertical gap $x$. These can be directly compared to experimental results.

\begin{figure}[t!]
\centering
\includegraphics[width=6.8cm]{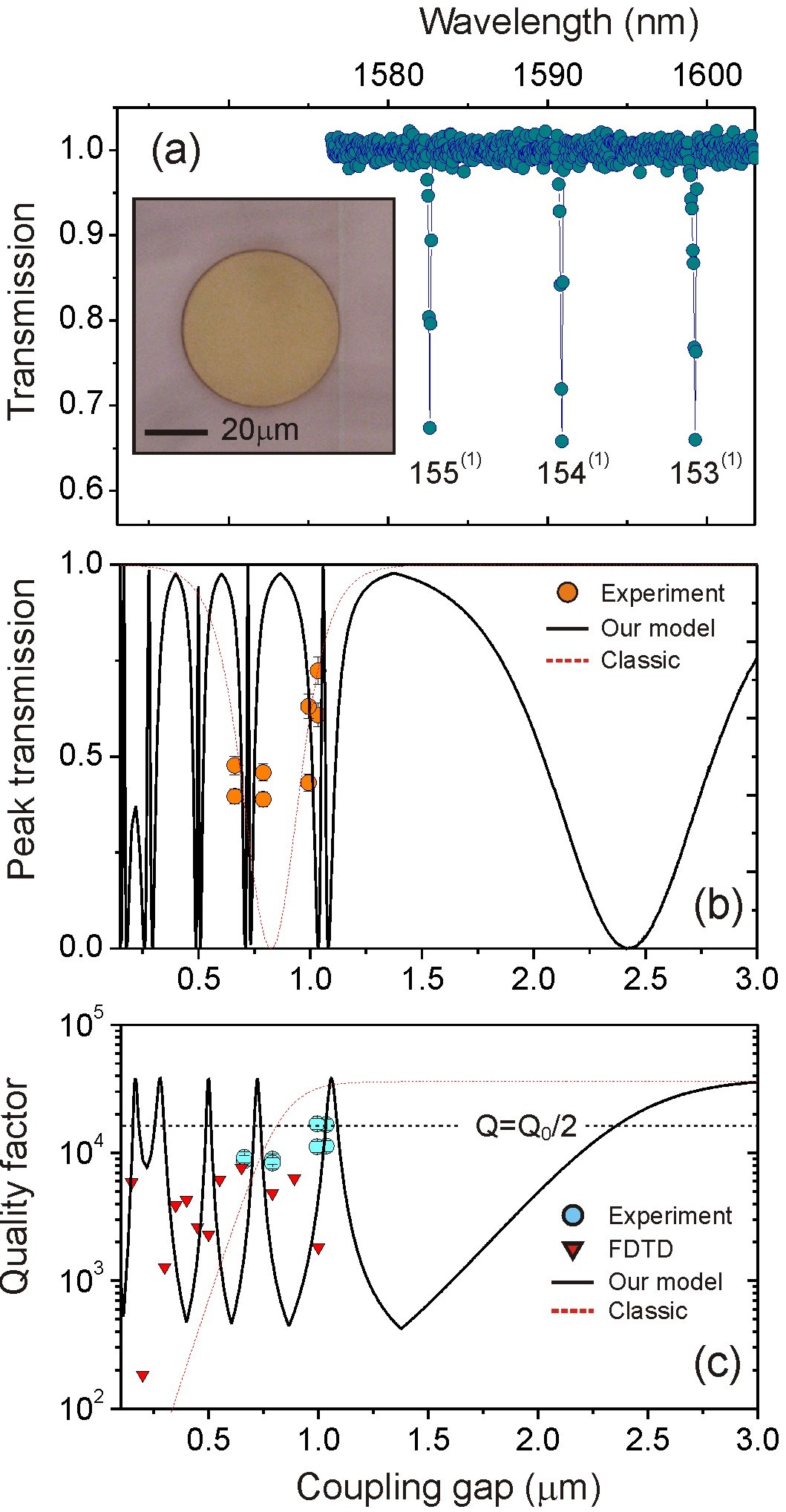}
\caption{(Color online) (a) An experimental transmission spectrum of a 50~$\mu$m WGM resonator, where some first radial family modes with different azimuthal mode numbers $M$ (labeled as $M^{(1)}$) are observed. Inset, optical image of the resonator and the buried waveguide. (b) The oscillatory behavior of the calculated (Eq.~\ref{tr1}) and the measured peak transmission (points). The dashed line is the result of the classical in-plane coupling theory. (c) The calculated $Q(x)$-factor (Eq.~\ref{gamma}, line), the measured Q-factor (full circles), the FDTD results (triangles) and the Q-factor obtained from the classical in-plane model (dashed line). The horizontal dotted line refers to half of the intrinsic Q-factor.} \label{tandq}
\end{figure}

In our experiments, 50~$\mu$m-diameter SiN$_x$ planar WGM resonators were fabricated and coupled vertically to single-mode SiN$_x$ waveguides as shown in Fig.~\ref{tandq}(a) (the only variable among different samples is the vertical gap thickness $x$). We refer to \cite{i3eptl} for a more detailed discussion of fabrication and characterization techniques.

From waveguide transmission experiments we have measured $T(\omega_0,x)$ and $Q(\omega_0,x)$ which are plotted as data points in Fig. \ref{tandq} (b) and (c), respectively. Then, we applied the developed theory to model the experimental results of several first radial family modes at $\lambda\approx1.59~\mu$m. Figure ~\ref{tandq}(b) shows that in fact, more than one minimum manifests in the waveguide transmission when the gap is monotonically reduced to zero. This result is in contrast with the in-plane case, where both the theory and experiments show a minimum at a unique critical coupling gap \cite{yariv,spillane,integr1,integr2,integr3,integr4}. However, this result is in agreement with the hypothesis of the ``flat zone/coupling length" interplay. The computed $Q(\omega_0,x)$, in its turn, shows multiple peaks with decreasing vertical distance (Fig.~\ref{tandq}(c)), while in the in-plane coupling scheme one observes a monotonic spoiling of the $Q$ because of the exponentially increasing coupling loss. The computed values are compatible with the experimental data both for $T$ and for $Q$-factor, though the experimental points are not numerous. It is interesting to note that the proposed model does not require an intermediate evaluation of the coupling constant and that by varying $\alpha$ one fits $T(\omega_0,x)$ and $Q(\omega_0,x)$ simultaneously. Thus, by fitting the experimental data with Eqs.~\ref{tr1} and \ref{gamma}, an intrinsic loss of the resonator $\alpha\sim2$~cm$^{-1}$ is estimated, which gives a reasonable intrinsic $Q_0$ of 38,000.

The agreement between the model and the data is even more evident when we compare the experimental data with the results of the in-plane model. We note here, that, within this model \cite{spillane}, $T(\omega_0,x)$ and the $Q(\omega_0,x)$ are indirectly related. In particular, using the experimental $T(\omega_0,x)$, one estimates the coupling constant $K=\frac{1\mp\sqrt T}{1\pm\sqrt T}$ and, then, fits $K$ with an exponential function of the gap. This procedure requires an estimation of the critical coupling gap $x_c$ for which $K\simeq1$, in order to switch between ``$\mp$" and ``$\pm$" for the undercoupled and overcoupled regimes. Afterwards, the mode quality factor is calculated as $Q(\omega_0,x)=Q_0/(1+K(\omega_0,x))$. In Figs.~\ref{tandq}(b) and (c) we report $T(\omega_0,x)$ and $Q(\omega_0,x)$, obtained using the in-plane model, which however do not provide a satisfactory fit to either experimental trends. Though the $T(\omega_0,x)$ might agree with the experimental data, $Q(\omega_),x)$ is markedly different.

Finally, we have performed full 3D finite-difference time-domain (FDTD) simulations of the fabricated system. We note here, that due to the fact that the simulations run in time-domain, the eigenmode linewidths have been extracted employing a harmonic inversion method, which essentially consists in fitting the cavity ring-down signal with a trigonometric decay function. While this algorithm is much robust with respect to a simple spectroscopic analysis of the transmitted field, it suffers an intrinsic limitation of numerical methods, i.e. the errors can be determined by the finite resolution and time-steps of the performed simulations. The FDTD results are also reported in Fig. \ref{tandq} (c) and show that at very small $x$ the cavity modes have from two to three orders of magnitude larger $Q$'s than those estimated from the in-plane model.

In conclusion, we have shown that in a vertical coupling geometry between a WGM resonator and a bus waveguide, a series of critical coupling points appear at several values of the vertical gap. This phenomenon originates from modal interference in the spatially extended coupling zone. By using a simple analytical model of the waveguide-resonator coupling, we have predicted an oscillating behavior between the undercoupling and overcoupling regimes as a function of the separation gap, which manifests both in the contrast and the linewidth of the dips in the waveguide transmission spectrum. The model provides a good agreement with experimental data and is further confirmed by numerical simulations. Our findings reveal an important and unexpected difference between vertical and in-plane coupling in optical circuits and can impact the design of photonic devices that are presently of high interest for both fundamental science \cite{hafezi} and for technological applications \cite{ghent2,nonlin-hydex}.

The authors wish to thank M.R. Vanacharla and N. Prtljaga for stimulating discussions. This work has been supported by NAoMI-FUPAT.


\begin{thebibliography}{99}
\bibitem{param}
The set of parameters (refractive index, $n$, and cross-sectional dimensions, $d$, $w$) used in the calculation correspond to the experimental ones and are $n_1=2.17$, $d_1=0.25\mu$m, $w_1=2.5\mu$m for the actual waveguide. The resonator parameters are $n_2=2.0$, $d_2=0.35\mu$m, and a radius of $r=25\mu$m. The coupling medium is silica with $n=1.456$.
\end{thebibliography}
\end{document}